\documentclass[prl,aps,showpacs,twocolumn,unsortedaddress]{revtex4-1}
\pdfoutput=1
\usepackage{graphics, bm}
\usepackage{psfrag}
\usepackage{amsmath}
\usepackage{amssymb}
\usepackage{epsfig}
\usepackage{subfigure}
\usepackage{grffile}
\newcommand{\beq}    {\begin{equation}}
\newcommand{\enq}    {\end{equation}}
\newcommand{\ceq}[1] {(\ref{#1})}

\newcommand{\rr}{{\bf r}}
\newcommand{\qq}{{\bf q}}

\newcommand{\df}     {\equiv}

\newcommand{\nimp}   {n_{\rm imp}}

\newcommand{\nrms}   {n^{(\rm rms)}}

\newcommand{\varGav} {\langle(\delta G)^2\rangle}
\newcommand{\sav}    {\langle\sigma\rangle}
\newcommand{\Gav}    {\langle G\rangle}

\newcommand{\ucfm}   {\varGav_{\rm ucf}}
\usepackage[usenames,dvipsnames]{color}			% Only used for author comments

\begin{document}

\title{Universal conductance fluctuations in Dirac materials in the presence of long-range disorder}

\author{E. Rossi$^1$, J. H. Bardarson$^{2,3}$, M. S. Fuhrer$^{4}$, S. Das Sarma$^{5}$}
\affiliation{
             $^1$Department of Physics, College of William and Mary, Williamsburg, VA 23187, USA\\
             $^2$Department of Physics, University of California, Berkeley, CA 94720, USA\\
             $^3$Materials Sciences Division, Lawrence Berkeley National Laboratory, Berkeley, CA 94720, USA\\
             $^4$CNAM, Department of Physics, University of Maryland, College Park, MD 20742-4111, USA\\
             $^5$CMTC, Department of Physics, University of Maryland, College Park, MD 20742-4111, USA
            }
\date{\today}

%-------------------------------------------------------
   
\begin{abstract}
We study quantum transport in Dirac materials with a single fermionic Dirac cone (strong topological insulators and graphene in the absence of intervalley coupling)
in the presence of non-Gaussian long-range disorder. We show, by directly calculating numerically the conductance fluctuations, that in the limit of very large system size and
disorder strength, quantum transport becomes universal. However, a systematic deviation away from universality is obtained for realistic system parameters.
By comparing our results to existing experimental data on $1/f$ noise, we suggest that many of the graphene samples studied to date are in a non-universal
crossover regime of conductance fluctuations.

\end{abstract}

%-------------------------------------------------------

\maketitle
With the successful isolation of graphene~\cite{novoselov2004}, and the discovery of topological insulators (TI), the study of {\it Dirac materials} has come to the fore. In a
Dirac material, the conduction and valence bands touch at an isolated set of points, called Dirac points (DP) (or {\it valleys}). The energy spectrum close to the DPs
is linear and the low energy properties of these materials are well described by a set of massless 2D Dirac fermions. Graphene is intrinsically a 2D
material~\cite{dassarma2011}, while in the case of a 3D topological insulator the Dirac fermions are realized as surface states whose presence is demanded by the non trivial
topology of the bulk energy bands~\cite{hasan2010,*hasan2010b,*qi2010b}. 
In this Letter, we report an exact numerical theoretical study of
quantum conductance fluctuations (CF) in the presence of non-Gaussian long-range
disorder in 2D Dirac materials, which should apply directly to both graphene (in the absence of intervalley scattering) and TI surface states.
Specifically by (non-Gaussian) long-range disorder we  refer to disorder whose spatial correlation
function decays algebraically. 
In {\em conventional}
parabolic band 2D electron liquids, taking into account screening,
the disorder potential is short-range even when due to charge impurities. In Dirac materials, in contrast,
due to the massless nature of the fermionic excitations, 
the disorder potential due to charge impurities retains
its long-range nature even when screening is taken into account;
in particular the spatial correlation function decays as $1/r^3$ at large distances.
As a consequence our work is the first to address the very fundamental and broad question
of how the long-range nature of the disorder affects the quantum transport
properties of 2D electron systems.

In graphene the
dominant source of disorder is believed to be from remote charge impurities~\cite{dassarma2011} (though other scattering mechanisms might also be of
relevance in certain cases~\cite{katsnelson2008, *guinea2008b, *wehling2010, *ferreira2011}). This type of disorder does not couple the two valleys of graphene and to a
good approximation graphene can be described by a single Dirac fermion. At the same time, a {\it strong} TI is characterized by an odd number of Dirac
cones in the surface spectrum (as opposed to a {\it weak} TI which has an even number). The dominant physics is captured by the case of a single Dirac fermion
which is experimentally realized in {\it e.~g.\ } Bi$_2$Se$_3$. In both cases the Hamiltonian is invariant under a time reversal symmetry $\mathcal{T}$
with $\mathcal{T}^2 = -1$, placing these systems in the symplectic (AII) symmetry class~\cite{altland1997}. Below we focus on the single Dirac fermion case. 

In the clean case, the density of
states goes to zero and transport is by quantum tunneling of evanescent modes rather than diffusion of propagating modes. This regime is 
referred to as {\it pseudo-diffusive}~\cite{tworzydlo2006}. 
Disorder drives
the system away from the pseudo-diffusive regime and into a symplectic metal phase characterized by weak
anti-localization~\cite{bardarson2007} (enhanced conductivity due to
destructive interference of time reversal symmetric paths~\cite{lee1985a}). 

These conclusions were reached by studying the effects of short range disorder. 
However in a Dirac material 
the disorder potential created by charge impurities retains its long range character even after screening 
has been taken into account \cite{nomura2006,*ando2006,*hwang2006b,*cheianov2006,*trushin2007,*trushin2008}. 
This leads to strong
carrier density inhomogeneities and, close to the DP
it induces {\it puddles} of electrons and holes. 
In this regime, conventional analytical methods fail and one needs to rely on numerical simulations.
This puddle formation due to random charged impurities in the environment has been experimentally observed both in graphene
\cite{martin2008,*zhang2009,*deshpande2009} and TIs \cite{Beidenkopf2011}.

In this paper 
we study the problem of quantum transport in the presence of long range disorder both in the puddle regime and away from the DP at large carrier
density, $n$. We study both quantum corrections to the average conductivity ($\sigma$)
[weak anti-localization (WAL)] and the CF arising from changes in interference patterns as external parameters are varied. In
conventional semiconductors and metals the amplitude of these fluctuations is a universal number giving rise to {\it universal conductance fluctuations} (UCF).
In terms of the scaling function $\beta(\sigma) = d\ln \sigma/d\ln L$, where $L$ is the sample length, 
the WAL is the first order term in the expansion in $1/\sigma$. It
happens that the second and third order terms are identically zero in the symplectic class~\cite{hikami1992} and therfore the WAL can be obtained already
for rather small values of $\sigma$. In contrast, higher order terms in the UCF are expected to be nonzero making the numerical calculation of UCF more
sensitive to finite size effects.
Moreover, in general,
the calculation of CF requires a much larger number 
of disorder realizations than the calculation of $\sigma$.
These facts make the numerical calculation
of CF in the symplectic class much more challenging than the calculation of $\sigma$ 
and WAL corrections.
Earlier numerical studies \cite{rycerz2007} of
UCF in graphene using tight binding models with short range disorder
found results consistent with the UCF theory
\cite{kechedzhi2008,*kharitonov2008}, however, even for short-range disorder, no systematic study of density
dependence or deviations from universality was attempted.

We demonstrate that just as in the case of
short range Gaussian disorder, long range disorder drives the system into the symplectic metal phase. However, the crossover regime before the universal
quantum transport sets in is considerably larger than in the case of short range disorder. This gives rise to CF with an amplitude which is smaller
than the UCF close to the DP but considerably larger at high $n$ which, as we argue, can lead to an intriguing non-monotonic behavior of the $1/f$ noise, which may have already been observed in graphene.
%--
We emphasize that our study is performed at zero temperature, in the absence of
inelastic scattering, conditions in which the phase coherence length $l_{\rm ph}$ is infinite.
Our analysis is valid as long as $l_{\rm ph}>L$.
The main goal of our work is to study the effect of the long-range
nature of the disorder on the universal properties of quantum
transport, for this reason we must have $l_{\rm ph}>L$ otherwise sample-dependent effects due to finite
$l_{\rm ph}$ would mask the sought after effects.
Moreover, any quantum transport measurement strives to be in this regime
that appears also to be the relevant one for current
experiments on mesoscopic samples of graphene. 
%----

To obtain the conductance $G$ of a single Dirac fermion we solve, using the transfer matrix method described in Ref.~\cite{bardarson2007}, the scattering
problem defined by the Hamiltonian:
$  H = v_F \mathbf{p}\cdot {\bm\sigma} + V_D(\rr) -\mu$.
Here $v_F$ is the Fermi velocity, ${\bm\sigma}$ the vector formed
by the Pauli matrices $(\sigma_x,\sigma_y)$ (in spin space for TIs and in sublattice space for graphene), $\mu$ the chemical potential which controls $n$, 
and $V_D$ is the disorder potential. We take a sample of length $L$ and width $W$ with periodic boundary conditions in the transverse direction. 
The solution of the scattering problem returns the transmission amplitudes $t$
that in turn give us the two-terminal conductance $G=R^{-1}=(g_s g_v e^2/h){\rm Tr}(t^\dagger t)$,
$R$ being the resistance and $g_s$ and $g_v$ the spin and valley
degeneracies respectively ($g_sg_v=4$ for graphene and $g_sg_v=1$ for
the surface of a TI). 
In the remainder for concreteness we set $g_sg_v=4$.
The conductivity is obtained via the relation $\sigma = (W dR/dL)^{-1}$ which minimizes the contribution of contact
resistance~\cite{adam2008d}. The CF are in turn obtained as the variance $\langle (\delta G)^2\rangle = \langle G^2 \rangle - \langle G
\rangle^2$, where the angle brackets denote the average over disorder realizations.

The scattering potential $V_D$ is induced by the remote impurity charges.
In the absence of screening, in
momentum space, $V_D(\qq) =(e^2/\epsilon_0) A(\qq)e^{-qd}/q$,
where $\epsilon_0$ is the background dielectric constant,
$d$ the average distance of the impurities from the graphene layer, and
$A(\qq)$ are random numbers with Gaussian distribution such that 
$\langle A\rangle=0$ and $\langle A^2\rangle=\nimp$, with $\nimp$ the impurity density.
The bare disorder is renormalized by the interactions replacing the dielectric constant $\epsilon_0$ by a dielectric function $\epsilon_0 \rightarrow
\epsilon(\qq)$. To calculate $\epsilon(\qq)$ in the presence of charge impurities we use the Thomas Fermi Dirac Theory (TFD) 
in which both Hartree and exchange-correlation
terms are taken into account \cite{rossi2008,rossi2010}.
The TFD approach is valid when the inequality $|\nabla n|/|n| < k_F$ is
satisfied. 
Close to the DP, due to the disorder, the root mean square of $n$ ($\nrms$) is much
larger than its average.
In this situation $\nrms$ should be used instead of $n$ in the inequality
above to verify the validity of the TFD theory \cite{rossi2008,rossi2009}.
Because $\nrms\approx\nimp$ \cite{adam2007,rossi2008} the
TFD theory is valid as long as
$\nimp$ is not too small ($\gtrsim 10^{11}{\rm cm}^{-2}$) \cite{brey2009,rossi2009}.
The TFD gives results for the carrier density profile $n(\rr)$
and $\epsilon(\qq)$ that compare well with both density functional theory (DFT)
results \cite{polini2008} and experiments \cite{martin2008,*zhang2009,*deshpande2009}
and is the only method available, for large samples
and ensembles, that takes into account nonlinear screening effects
that dominate close to the DP.

For purposes of comparison it is useful to consider different models of screening. In particular one can turn off 
exchange-correlation terms in the TFD. We denote the corresponding dielectric function by $\epsilon_\text{nxc}$, the full TFD dielectric function being
$\epsilon_\text{TFD}$. The cases of doping dependent but uniform
screening $\epsilon^{-1} \rightarrow \epsilon_{dds}^{-1} = q/(q+4r_s k_F)$, 
where $r_s=e^2/(\epsilon_0\hbar v_F)$,
and constant screening $\epsilon^{-1} \rightarrow \epsilon_{cs}^{-1} =
q/(q+q_s)$ will also be considered. Finally, we also compare with the case of Gaussian correlated disorder defined by $\langle V_D(\rr)V_D(\rr') \rangle =
K_0/(2\pi\xi^2)\exp(-|\rr-\rr'|^2/2\xi^2)$. $K_0$ is a dimensionless measure of the disorder strength and $\xi$ the correlation length. This is the type of disorder correlation considered in most prior numerics, and in analytical considerations
(where the limit $\xi \rightarrow 0$ is routinely taken).

The analytical theory of UCF requires diffusion, i.e. $L \gg\ell$,
where $\ell$ is the mean free path.
In this
limit one can perform a controlled diagrammatic calculation of transport properties. 
Adapting the results of the short-range calculation in \cite{kharitonov2008} to our geometry, this approach predicts a universal (independent of
microscopic parameters, such as $n_\text{imp}$, $K_0$ and $\xi$) value of the CF given by
\beq
 \ucfm = \left(\frac{e^2}{\pi^2h}\right)^2\sum_{\begin{subarray}{l} n_x=1\\ n_y=-\infty\end{subarray}}^{\infty}
   \frac{12 g_s g_v}{\left[n_x^2 + 4\left(\frac{L}{W}\right)^2 n_y^2\right]^{2}}.
 \label{eq:ucf}
\enq
\begin{figure}[!tb]
 \begin{center}
  \includegraphics[width=8.5cm]{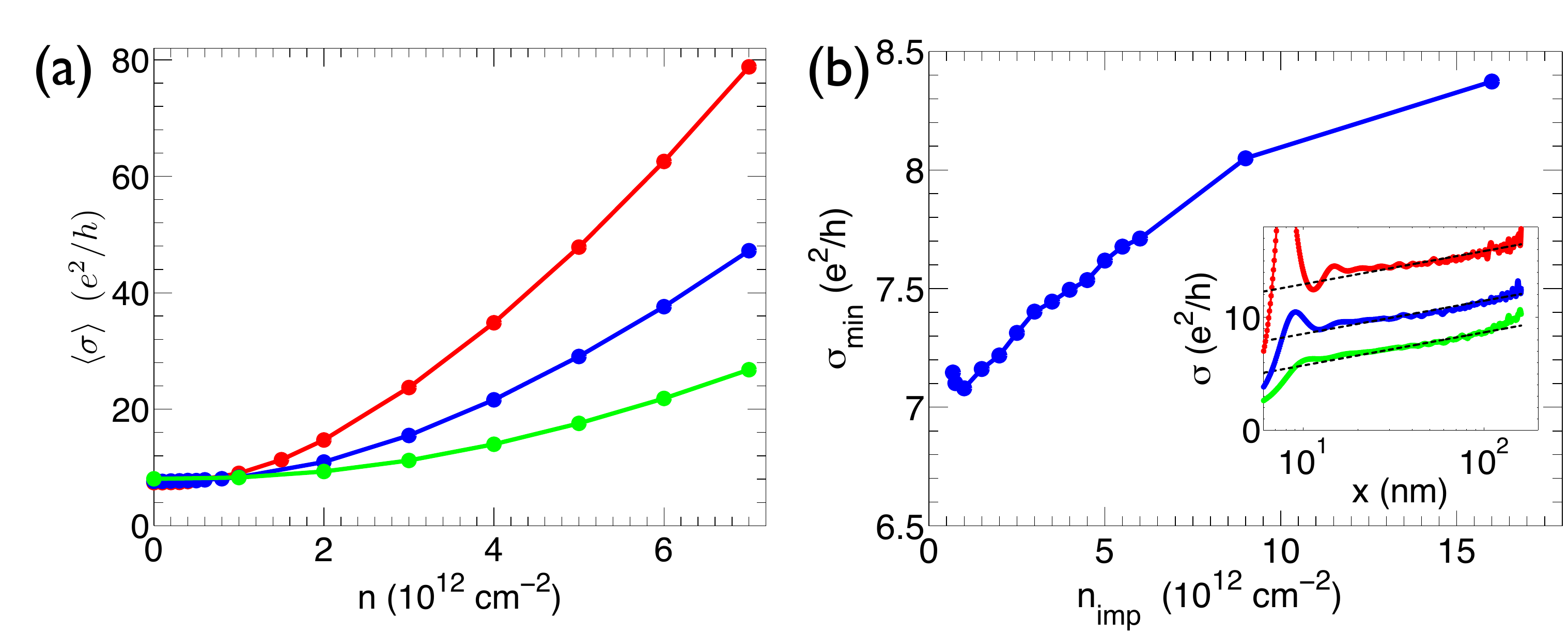} 
  \caption{
           (Color online). 
           (a) $\sav$ as a function of $n$ for $\nimp=(3, 5, 9)\times 10^{12}{\rm cm}^{-2}$ from top to bottom.
           (b) $\sav$ at the DP as a function of $\nimp$ for the case of Coulomb disorder. The inset shows
               $\sav$ as a function of sample length $L\df x$ for $W=160$~nm, and $n=(3, 2, 1)\times 10^{12}{\rm cm}^{-2}$
               from top to bottom.
               The dashed lines show $d\sigma/d\ln L=4e^2/\pi h$.
        } 
  \label{fig:sigma}
 \end{center}
\end{figure} 
\begin{figure}[!tb]
 \begin{center}
  \includegraphics[width=8.5cm]{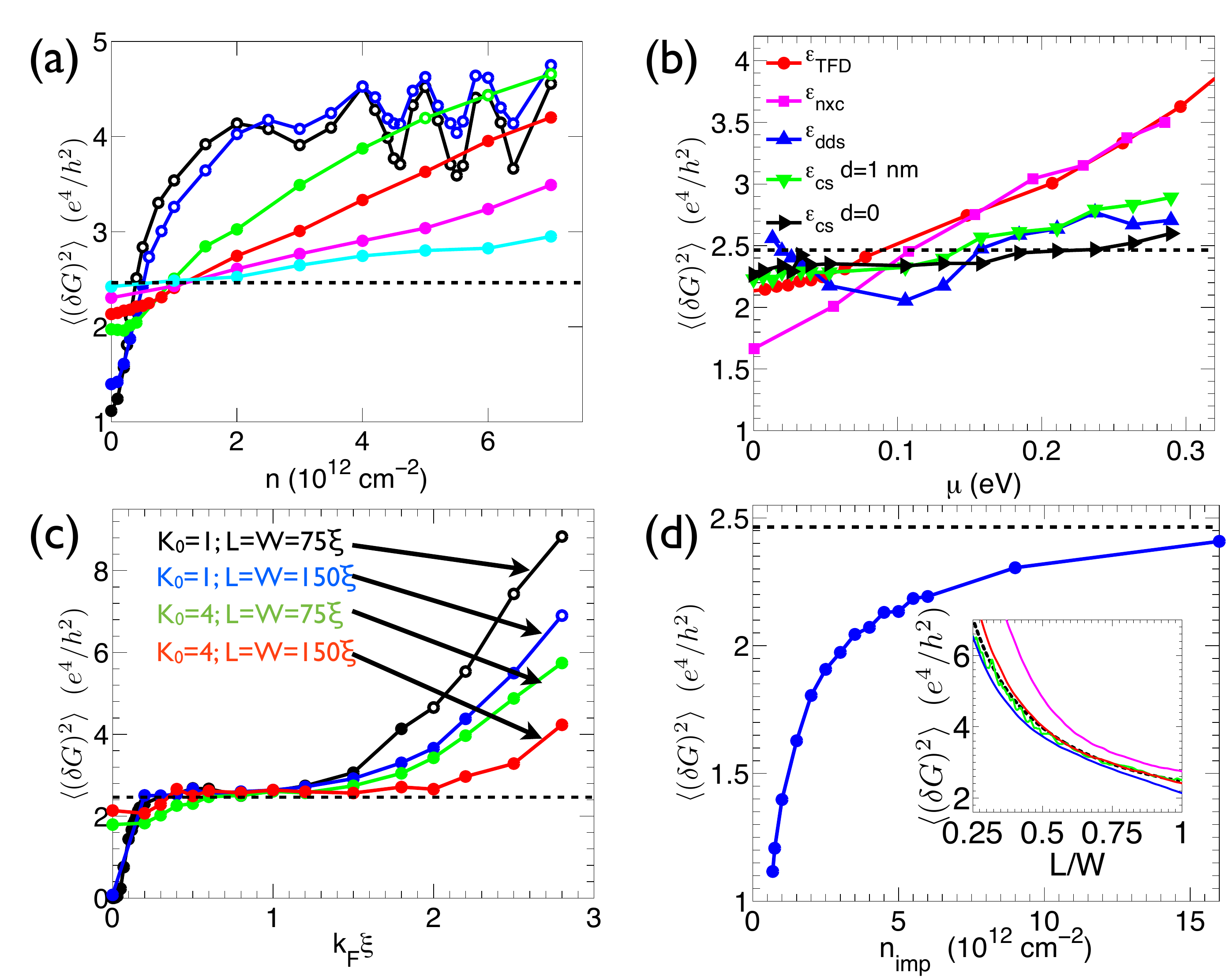} 
  \caption{
           (Color online). 
           (a) $\varGav$ as function of $n$ for the case of long-range disorder 
               ($L=160$~nm) and different values of $\nimp$;
               from top to bottom (for $n=2\times 10^{12}{\rm cm}^{-2}$):
               $\nimp=(0.682, 1, 3, 5, 9, 16)\times 10^{12}{\rm cm}^{-2}$.
	   (b) $\varGav$ as a function of $\mu$ for different types of disorder potential (see text).
           (c) $\varGav$ as a function of $k_F\xi$ for Gaussian disorder.
           (d) $\varGav$ at the DP as a function of $\nimp$ for the case of long-range disorder
               ($L=160$~nm).
               In the inset 
               $\varGav$ as a function $L/W$;
               the top line shows the results for 
               $\nimp=5\times 10^{12}\;{\rm cm}^{-2}$ and 
               $n=3\times 10^{12}\;{\rm cm}^{-2}$; the solid line below,
               and the bottom one, show the results for the same $\nimp$
               and $n=10^{12}\;{\rm cm}^{-2}$ and $n=0$ respectively.
               Between these two lines there is the line showing the results
               for the case of Gaussian disorder with $K_0=1$ and
               $k_F\xi=0.8$.
               The dashed lines in all the panels show the universal value of Eq.~\ceq{eq:ucf}
         } 
  \label{fig:deltaG}
 \end{center}
\end{figure} 

Fig.~\ref{fig:sigma}(a) shows the dependence of 
$\sav$ on the doping $n$ for three different values of $\nimp$. $\sav$ approaches linear dependence on $n$ at large
densities in qualitative agreement with previous theoretical results 
\cite{hwang2006c,rossi2009} and experiments \cite{tan2007,*chen2008}. At the DP, the conductivity ($\sigma_{\rm min}$) is enhanced
by disorder, see Fig.~\ref{fig:sigma}(b), in contrast to what is obtained in Boltzmann type of theories extended to the DP, but in agreement with simulations with
short range disorder. For large system sizes and disorder strengths $\sigma$ increases logarithmically with system size, as shown in the inset of Fig.~\ref{fig:sigma}(b), 
consistent with weak anti-localization which predicts $\sigma \sim (g_sg_v/\pi) \ln L$~\cite{hikami1980}.  

Fig.~\ref{fig:deltaG}(a) shows $\varGav$
as a function of $n$ for several values of $\nimp$.
The dashed lines in Fig.~\ref{fig:deltaG} show the universal value of Eq.~\eqref{eq:ucf}.
We observe that close to the DP the CF tend to be smaller than the UCF value, while at large $n$ they are larger. As
$\nimp$ is increased, the CF approach the universal value with the set of curves crossing roughly at the UCF value.
Similar trends are observed as a function of system size. This suggests that in the thermodynamic limit the CF in the presence of long range disorder, approach
the universal value {\it independent} of $n$. In this
limit the diffusive symplectic metal is an accurate description of the
system. In contrast, percolation predicts CF larger than UCF at
the DP \cite{cheianov2007}.

To understand better the deviation from universality and to quantify the 
behavior observed in Fig.~\ref{fig:deltaG} we estimate $\ell$
using the Boltzmann theory, which gives $\ell =f(r_s, k_F d)\sqrt{n}/\nimp$, \cite{nomura2006,*ando2006,*hwang2006b,*cheianov2006}, where 
$f$ is a function that for $d\lesssim 1$~nm
depends very weakly on $k_F d$ 
\cite{hwang2006c,adam2008b}. For $d=1$~nm and $r_s=0.8$, $f\approx 5.66$ almost independent of $k_F$.
In Fig.~\ref{fig:deltaG} we represent with open symbols those data points for which $\ell > 0.25L$, revealing that the observed oscillations at large
$n$ and small $n_\text{imp}$ are ballistic effects. It also suggests that the departure of the CF from the UCF value at high
$n$ is due to the fact that we are in the ballistic to diffusive
crossover regime. 
Similarly, close to the DP, the deviations are attributed to a
pseudo-diffusive to diffusive crossover (the estimates of $\ell$ are not as reliable in this regime).

To better understand the role played by the disorder-induced density inhomogeneities we plot in Fig.~\ref{fig:deltaG}(b) the CF
for the different variants of screening discussed above. We observe that neglecting exchange-correlation terms in the TFD
increases the deviation of the CF from the universal value close to the DP. 
This is due to the fact that in the absence of exchange-correlation the amplitude of the 
density fluctuations are increased \cite{rossi2008}. 
In contrast, assuming
uniform screening, either doping dependent or constant, gives results that are closer to the UCF value.% for smaller values of $\nimp$. 
The results of Fig.~\ref{fig:deltaG}(b) clearly show that the presence of strong carrier
density inhomogeneities increases the range of $n$ and $\nimp$ for which the
transport is in the crossover regime for which the CF
differ from the universal value.

In Fig.~\ref{fig:deltaG}(c) we show the CF for the gaussian correlated potential as a function of $k_F\xi = (4\pi n/g_sg_v)^{1/2}$.
In this case from the Boltzmann theory we have $\ell=2\sqrt{2\pi}\xi^3 k_F^2/K_0$.
As in the case of charge impurities we 
find that in the pseudo-diffusive regime $\varGav$ is smaller than the UCF value, and larger in the ballistic high density regime, but approaches
the UCF value as the impurity strength $K_0$ or system size $L$ are increased. In contrast to the long range case, there is a large range of intermediate
values of $n$ where the CF quickly goes to the UCF value.
Thus, the crossover regime is strongly suppressed for short-range
disorder compared with long-range disorder.
In the presence of spatial correlations among charge impurities the 
long-range nature of the disorder is suppressed compared to the case
of uncorrelated impurities \cite{li2011,*yan2011} and so we predict that
the crossover regime  will also be reduced.

The trend that emerges from Fig.~\ref{fig:deltaG}(a) is that as $\nimp$ increases the 
CF approach the universal value. This is further demonstrated in
Fig.~\ref{fig:deltaG}(d) which shows the value of the CF at the DP as a function of $\nimp$ for the case of charge impurities.
We see that for very large $\nimp$, for which $\ell/L<<1$, the CF saturate to the universal value (dashed line).
In the inset to Fig.~\ref{fig:deltaG}(d) we compare the aspect ratio dependence of the CF to the analytical expression~\eqref{eq:ucf}. Both in the case of long range
and Gaussian disorder the curves agree well within the universal regime. In the crossover regime the CF follow the same trend but
with an amplitude that differs from the UCF prediction.

Taken together these results suggest that even in the presence of long-range disorder, the system is eventually (large enough $L$ and/or disorder
strength) driven into the universal symplectic metal
phase and that, however, the crossover regime in which neither ballistic or universal diffusive physics is applicable is 
very large in the presence of long-range Coulomb disorder and puddles.

To connect these results with experiments we consider the relevance of the CF results to the $1/f$ noise.
In 2D (and 1D) the displacement of a single
defect can cause a change in $G$ of the order of $e^2/h$ \cite{altshuler1985c,*feng1986,*birge1989}.
One of the probable sources of resistance fluctuations is thermally
activated motion of defects. Assuming that the defects move on
a time scale $\tau\gg \tau_{\rm inelastic}$ and that the hopping distance
is uncorrelated with $\tau$, for the spectrum of the resistance
fluctuations $S_R(\omega)=\int dt \langle \delta R(t)\delta R(0)\rangle e^{i\omega t}$ we have~\cite{feng1986}
\begin{equation}
 S_R(\omega) = R^2\left[ \frac{\varGav}{\Gav^2}\right]\int\frac{2\tau}{\omega^2\tau^2 + 1}P(\tau)d\tau
\end{equation}
where $P(\tau)$ is the probability distribution of the time that it takes a defect to move.
If the motion of defects is thermally activated then $P(\tau)$ is quite broad and 
$S_R(\omega)$ will be approximately $1/f$.
We can then calculate the dependence with respect to the doping of the strength of the $1/f$ noise by calculating the coefficient
$\varGav/\Gav^2$, or equivalently $\varGav/\sav^2$.
\begin{figure}[!tb]
 \begin{center}
  \includegraphics[width=8.5cm]{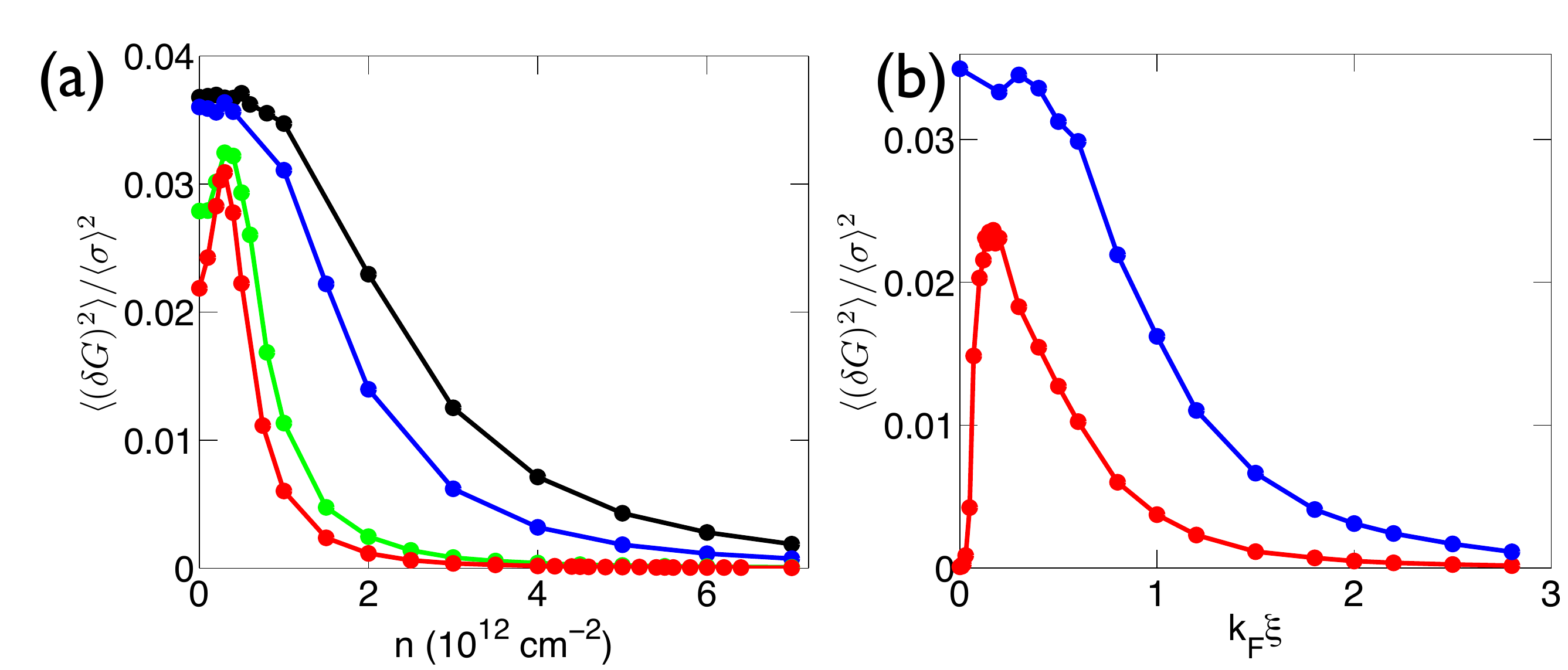} 
  \caption{
           (Color online).     
           (a) $\varGav/G^2$ as function of $n$ for the case of long-range disorder
               ($L=W=160$~nm) and different values of $\nimp$;
               from top to bottom (for $n=2\times 10^{12}{\rm cm}^{-2}$):
               $\nimp=(5, 3, 1, 0.682)\times 10^{12}{\rm cm}^{-2}$.
           (b) $\varGav/G^2$ as a function of $k_F\xi$ for the case of Gaussian disorder for $K_0=4$, top,
               and $K_0=1$, bottom. $L=W=75\xi$.
         } 
  \label{fig:VarGoverg}
 \end{center}
\end{figure} 

Fig.~\ref{fig:VarGoverg} shows  $\varGav/\sav^2$ as a function of $n$ 
for both charge disorder and Gaussian disorder. In both cases we see that at low disorder strengths $\varGav/\sav^2$
depends non-monotonically on $n$. The reason is that in the pseudo-diffusive regime 
$\varGav$ grows rapidly with $n$ while $G$ grows
very slowly, whereas at larger values of $n$ the transport is diffusive and $\varGav$
is almost constant while $G$ grows rapidly with $n$. In contrast, in the symplectic metal regime the ratio $\varGav/\sav^2$ decreases
monotonically with $n$.  
We conclude that the non-monotonic dependence
of the $1/f$ noise is a generic property of the crossover
regime between pseudo-diffusive and diffusive regimes.
The $1/f$ noise of graphene has been measured experimentally 
\cite{xu2010,*pal2011} and its dependence on $n$ has been shown
to qualitatively follow the non-monotonic behavior shown in Fig.~\ref{fig:VarGoverg}.
This suggests that the samples in these experiments are not fully in the diffusive regime but rather in a non-universal crossover regime.

In conclusion, we have calculated the amplitude of the conductance fluctuations 
in Dirac materials with a single Dirac cone and in the presence of disorder with various range and screening properties.
Our work is the first to consider the effect of the long-range nature of disorder on
the quantum transport properties of 2D electron systems.
In the limit that the mean free
path is much smaller than the system size, the CF approach the universal value predicted by diagrammatic calculations. Before reaching the universal
value the CF are systematically smaller than the universal value close to the DP but larger away from it. 
In particular, for system parameters realistic to graphene, the CF
seems to deviate from the universal value leading to a non-monotonic dependence of $1/f$ noise, as recently observed in
experiments~\cite{xu2010,*pal2011}.

It is a pleasure to acknowledge Piet W. Brouwer for very helpful
discussions. This research was supported by the Jeffress Memorial
Trust, Grant No. J-1033 (ER), DOE DE-AC02-05CH11231 (JHB),
ONR-MURI (SDS, MSF), and NRI-SWAN (SDS). 
Computations were carried out on the
SciClone Cluster at the College of William and Mary and
University of Maryland High Performance Computing Cluster (HPCC).

%-------------------------------------------------------
%-------------------------------------------------------

%-------------------------

\end{document}